\documentstyle[aps,pre,epsfig]{revtex}

\begin{document}
\draft

\title{{\bf Equation of state of a strongly magnetized hydrogen plasma}}
\author{M.Steinberg, J. Ortner, and W.Ebeling}
\address{{\it Institut f\"ur Physik, Humboldt Universit\"{a}t zu Berlin, 
Invalidenstr. 110, D-10115 Berlin, Germany}}

\date{\today}
\maketitle

\begin{abstract}
The influence of a constant uniform magnetic field on the thermodynamic properties of a partially ionized hydrogen plasma is studied. Using the method of Green's function various interaction contributions to the thermodynamic functions are calculated. The equation of state of a quantum magnetized plasma is presented within the framework of a low density expansion up to the order $e^4 \, n^2$ and, additionally, including ladder type contributions via the bound states in the case of strong magnetic fields ($ 2.35 \times 10^{5} T \ll B \le 2.35 \times 10^{9} T$). We show that for high densities ($n \approx 10^{27-30} m^{-3}$) and temperatures $T \approx 10^5 - 10^6 K$ typical for the surface of neutron stars nonideality effects as ,e.g., Debye screening must be taken into account. 
\end{abstract}

\pacs{52.25.Kn, 05.70.Ce, 97.60.Jd}

\section{Introduction}
The calculation of the equation of state (EOS) of a multi-component quantum plasma consisting of charged particles interacting via the Coloumb potential is of theoretical interest as well as of practical relevance, e.g., for astrophysical systems such as stars. The aim of this paper is to derive a low-density expansion for the equation of state (EOS) of a two-component plasma embedded in an external constant magnetic field. This problem was recently tackled by Cornu \cite{Cornu} and Boose\&Perez \cite{Boose&Perez} who derived a formally exact virial expansion of the EOS by using a formalism which is based on the Feynman-Kac path-integral representation of the grand-canonical potential. 

In this paper we will employ the method of Green's function. As the calculations are carried out for a nonrelativistic quantum system, we restrict ourselves to magnetic field strengths $B<B_{rel}$ which is given by $B_{rel}=m_e^2 c^2/(e\hbar)\approx 4.4 \times 10^9 T$.Further we will use  an expansion of the magnetized plasma pressure in terms of the fugacity $z=e^{\beta \mu} $ to obtain the EOS of a weakly coupled magnetized plasma. Thus we can derive explicit expressions for various contributions to the quantum second virial coefficient. Though the formalism is formally valid only for low densities the obtained explicit expressions are appropriate even at sufficient high densities as the magnetic field increases the domain of classical behavior towards higher densities. The second virial coefficient contains both scattering and bound state contributions of two-particle states. Being interested in the thermodynamic properties of quantum magnetized plasmas the influence of the magnetic field on the energy eigen states of a two-particle states has to be taken into account.

Usually the magnetic field is measured by the dimensionless parameter $\gamma = \hbar \omega_c/(2 Ry) = B/B_0 $ where $\hbar \omega_c$ is the electron cyclotron energy, $B_0\approx2.35\times 10^5 T$, and $Ry=e^2/(8\pi \epsilon_0 a_B) \approx 13.605 eV$ is the ionization energy of the field-free hydrogen atom. Whenever $\gamma > 1$, i.e. the cyclotron energy is larger than the typical Coulomb energy, the structure of the hydrogen atom is dramatically changed. This problem has been approached by several authors \cite{Lai&Salpeter,Wunner1,Potekhin,Liberman&Johansson}. Using the results of these authors we study the influence of bound and scattering states on thermodynamic properties of magnetized plasmas. 

Recently the problem of ionization equilibrium of hydrogen atoms in superstrong magnetic fields ($\gamma \gg 1$) was considered by Lai\&Salpeter\cite{Lai&Salpeter}. They proposed an ideal Saha equation of a hydrogen gas including bound states but neglecting screening effects and scattering contributions to the second virial coefficient. Using the EOS obtained in our derivation we construct a modified Saha equation which takes into account nonideality effects as well. \par
The paper is organized as follows. In section II, we discuss the method which is used to calculate thermodynamic functions and derive analytical results for the scattering contribution in section III. An approximate result for the bound state contributions is given in section IV and the equation of state is presented in section V. Finally, we use our results to derive a generalized Saha equation and compare the degree of ionization with the results of the ideal Saha equation in section VI.   
\section[Fugacity expansions of the thermodynamic functions]{Fugacity expansions of the thermodynamic functions}

We consider a two-component charge-symmetrical system of N spin half particles of charge (-e) and mass $m_e$ and N spin half particles of charge e and mass $m_i$. In general, the total pressure can be split into ideal contributions and interaction contributions 
\begin{equation}
\label{2.0} p=p_{id} + p_{int} \, .
\end{equation}
The pressure and the particle density of an ideal plasma in a homogeneous magnetic field ${\bf B}=(0,0,B_0)$ are given by a sum of Fermi integrals over all Landau levels n 
\begin{equation}
\label{2.1} p_{id} = kT \sum_a \, \frac{2x_a}{\Lambda_a^3} \, {\sum_{n=0}} ^\prime f_{\frac{1}{2}}(\ln z_n^a) \, \, , \hspace{1cm} n=\sum_{a} \frac{2x_a}{\Lambda_a^3} \, {\sum_{n=0}} ^\prime \, f_{-\frac{1}{2}}(\ln z_n^a)
\end{equation}
($x_a=\hbar \omega_c^a/(2kT)$ with $\omega_c^a=|e_a|B_0/m_a$, $\Lambda_a =h / \sqrt{2\pi m_akT}$, and $z_n^a=\exp{[\beta(\mu-n\hbar\omega_c^a)]}$). The prime indicates the double summation due to the spin degeneracy except for the $n=0$ level. \par
The interaction part of the pressure for sufficiently strong decaying potentials may be written in terms of a fugacity expansion 
\begin{equation}
\label{nd8} \beta  (p - p_{id}) = \sum_{ab} {\tilde z}_a {\tilde z}_b \, B_{ab} + \sum_{abc} {\tilde z}_a {\tilde z}_b {\tilde z}_c\, B_{abc} + ... \, ,
\end{equation} 
where we have introduced the modified fugacities
\begin{equation}
\label{nd6} {\tilde z}_a = z_a \, \frac{2}{\Lambda_a^3} \, \frac{x_a}{\tanh(x_a)} \, .
\end{equation}
In the limit of small densities we have ${\tilde z}_a \rightarrow n_a$. We focus on the calculation of the second virial coefficient $B_{ab}$ which is defined by
\begin{equation}
\label{nd6.44}  B_{ab} = \frac{1}{2 \Omega} \left( \frac{\Lambda_a^3}{2} \frac{\tanh(x_a)}{x_a} \right)   \left( \frac{\Lambda_b^3}{2} \frac{\tanh(x_b)}{x_b} \right)    \, {\bf Tr} \, (e^{-\beta \widehat H_{ab}^{\lambda=1}} -   e^{-\beta \widehat H_{ab}^{\lambda=0}}) \, ,
\end{equation} 
$\widehat H_{ab}^\lambda$ is the Hamiltonoperator of the two particle system with the interaction potential $V_{ab}({\bf r})$
\begin{equation}
\label{nd75} \widehat{H}_{ab}^\lambda = \left( \frac{({\bf p}_a-e_a{\bf A}_a)^2}{2m_a}+\mu_{B}^a B_0 \sigma_z \right) + \left( \frac{({\bf p}_b-e_b{\bf A}_b)^2}{2m_b}+\mu_{B}^b B_0 \sigma_z \right) +\lambda V_{ab}({\bf r}) \, , \hspace{1cm} \sigma_z=-1,+1
\end{equation} 
and $\widehat H_{ab}^{\lambda=0}$ of the noninteracting system. The additive term $\mu_{B}^a B_0 \sigma_z $ describes the coupling between the intrinsic magnetic moment ($\mu_{B}^a = e_a \hbar /(2m_a)$) of the charged particles and the magnetic field. However, in the case of particles interacting via the Coulomb potential $V_{ab}({\bf r})=e_a e_b/(4\pi\epsilon_0\mid {\bf r}_a - {\bf r}_b \mid)$ the second virial coefficient defined by Eqs.(\ref{nd6},\ref{nd75}) is divergent. In order to obtain a convergent expression one has to perform a screening procedure. Such a technique is well established in the zero magnetic field case \cite{Vedenov&Larkin,Kraeft&Kremp&Kilimann,EKK} and can be easily extended to the nonzero magnetic field case. This program was also carried out by Cornu \cite{Cornu} and Boose\&Perez \cite{Boose&Perez} who used the Feynman-Kac formalism to derive a virial expansion for a magnetized multi-component system. Using the methods as described in \cite{Vedenov&Larkin,Kraeft&Kremp&Kilimann,EKK} the convergent second virial coefficient of a plasma may be split into a scattering and bound state contribution. In contrast to the zero magnetic field case an exact calculation of the convergent second virial coefficient in terms of scattering phase shifts is very complicated. Therefore we will give a perturbation expansion of the scattering part in terms of the interaction parameter $e^2$ up to the order $e^4$ and use an approximate expression for the bound state part which is valid in the case of strong magnetic fields ($\gamma > 100$). We may employ the method of Green's function. The starting point is the observation that the equation of state is connected to the average interaction energy $\langle \lambda V_{ab} \rangle$ by a charging process 
\begin{equation}
\label{nd3}  p-p_{id} = -\frac{1}{\Omega} \, \int_0^1 \frac{d\lambda}{\lambda} \, \langle V_{ab} \rangle_\lambda  \, ,
\end{equation}
$\Omega$ is the volume of the system. Taking into account many body effects thermodynamic functions may be expressed by a screened potential $V_{ab}^s$. By this method the divergencies due to the long range Coulomb force are removed. Then the pressure is given by the equation 
\begin{equation}
\label{vgf6}  \beta \left( p-p_{id} \right) = \frac{1}{2\Omega} \, \sum_{ab} \int_0^1 \frac{d\lambda}{\lambda} \int  d1 \, d2 \, \left(V_{ab}(12\lambda) \, {\rm G_a(11) \, G_b(22) } + V_{ab}^s(12\lambda) \, \Pi_{ab}(121^{++}2^+ \lambda) \right)   \hspace{0.12cm}.
\end{equation}
Here the first term is the Hartree approximation given in terms of the free particle Green's function  $G_a(11)$ and $\Pi_{ab}$ denotes the polarization function. For low density systems it is necessary to calculate bound state contributions to the thermodynamic functions. Therefore we apply the ladder approximation for $\Pi_{ab}$
\begin{equation}
\label{nd8.1} \beta \left( p-p_{id} \right) =\frac{1}{2\Omega} \, \sum_{ab} \int_0^1 \frac{d\lambda}{\lambda}  \, \left( \hspace{2.6cm} \begin{minipage} [htbp]{2cm} \centerline{\psfig{figure=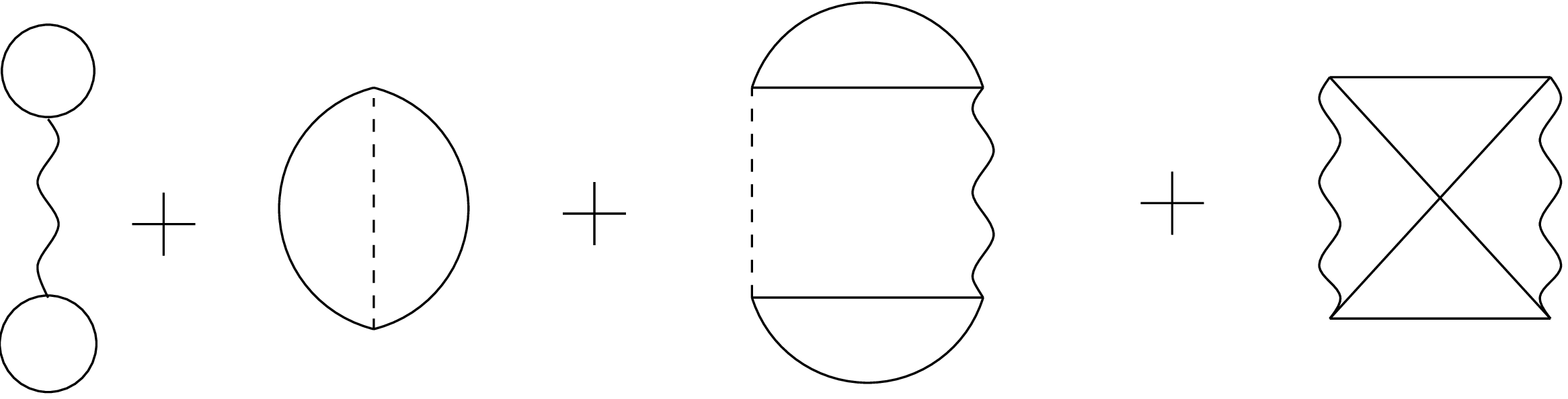,width=1.4cm,angle=-90}} \end{minipage} \hspace{2.6cm} \right)+\frac{1}{2\Omega} \, \sum_{ab} P_3 \, \int_0^1 \frac{d\lambda}{\lambda}  \, \left( \hspace{0.2cm} \begin{minipage} [htbp]{1cm} \centerline{\psfig{figure=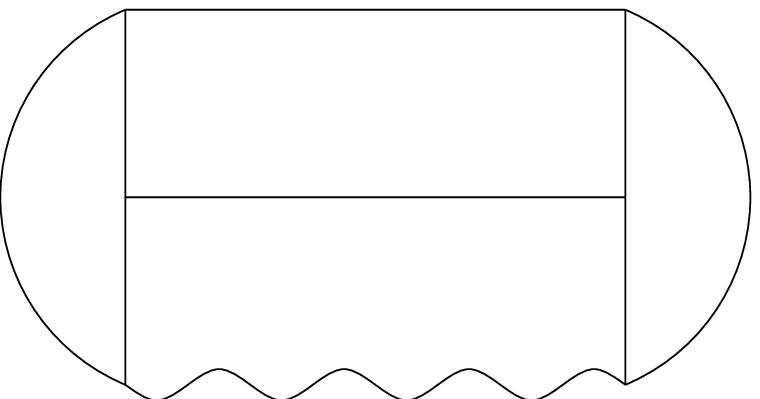,width=1.4cm,angle=-90}} \end{minipage} \hspace{0.2cm} \right)\, .
\end{equation}
To avoid double counting we have introduced the operator $P_3$ which subtracts contributions of the order $V_{ab}^s$ and $(V_{ab}^s)^2$.    
We may divide $p_{int}$ into a bound state contribution $p_{int}^{bound}$ and a scattering state contribution $p_{int}^{scatt}$
\begin{equation}
\label{nd7} p_{int}=p_{int}^{bound}+p_{int}^{scatt}. 
\end{equation}
In the case of a Coloumb potential this division is not trivial as the atomic partition function is divergent due to the infinite number of bound states at the continuum boundary. This problem has been extensively discussed in the zero magnetic field case \cite{EKK}. One can solve this problem in a natural way by introducing a renormalized sum of bound states 
\begin{equation}
\label{nd8.2} p_{int}^{bound} = {\tilde z}_e {\tilde z}_i P_3 \, B_{ab}^{bound} \, ,
\end{equation}
where at zero magnetic field $B_{ab}^{bound}$ is given by the Planck-Larkin partition function \cite{Larkin}. This division is somewhat arbitrary but guarantees the convergence of the bound state partition function even at vanishing magnetic field. We mention that this division does not affect the results of the thermodynamic potentials.

\section[scattering state contribution]{scattering state contribution}

We consider all diagramms up to the order $e^4$ in the interaction parameter. A diagrammatic representation of the perturbation expansion takes the form

\begin{equation}
\label{2.1.46} \hspace{0cm} \hspace{-1cm}{ \beta p_{int}^{scatt} =  \frac{1}{2\Omega} \, \sum_{ab} \int_0^1 \frac{d\lambda}{\lambda}}  \, \left( \hspace{2.8cm} \begin{minipage} [htbp]{2cm} \centerline{\psfig{figure=druck1.eps,width=1.4cm,angle=-90}} \end{minipage} \hspace{2.6cm} \right) \hspace{0.12cm}. \nonumber\\
\end{equation}
These diagrams are the Hartree term, the Montroll-Ward term, the Hartree-Fock term, and the exchange $e^4$ term, respectively. The solid lines represent the uncorrelated Green's function for a charged particle in a magnetic field \cite{Horing}. Hence our calculations are valid at arbitrary magnetic field strength. The divergence of the Montroll-Ward graph is avoided by introducing a screened potential line. The screened interaction potential $V^s$ is evaluated in the random phase approximation $V^s({\bf q},\omega)=V({\bf q})/(1-V({\bf q})\, \Pi^{RPA}({\bf q},\omega))$. At low densities $V^s$ can be approximated by a statically screened potential $V^s= e^2/(\epsilon_0 \, [q^2+\kappa^2])$ with $\kappa^2 = (e^2/\epsilon_0) \, \Pi^{RPA}(0,0)=\beta \, (e^2/\epsilon_0) \, \left({\tilde z}_e + {\tilde z}_i\right)$.
In the following calculations all results are obtained by setting the distribution function $f_0(\omega)=e^{\beta \mu}e^{-\beta \omega}$, i.e., in the nondegenerate limit $n \lambda^3 \, \tanh(x)/x\ll1$. The Hartree term vanishes due to the electroneutrality.
\subsection{GREEN'S FUNCTION FOR THE MAGNETIC FIELD PROBLEM}
In this section we represent the uncorrelated Green's function for a charged particle moving in a constant magnetic field in a closed form. The Green's function is the solution of the equation of motion (using symmetric gauge and setting $\hbar=1$):
\begin{equation}
\label{3114}\left( \frac{\triangle_{\bf R}}{2m}-\frac{m\omega_c^2}{8}(X^2+Y^2)+\frac{\omega_c}{4} {\widehat L}_z -\mu_B B\sigma_z+i\frac{\partial}{\partial T}\right) \, G^\prime({\bf R} , T)= \delta(R) \, \delta(T)  \hspace{0.12cm}.
\end{equation}
$G^\prime({\bf r,r^\prime},T)$ can be expressed in terms of the correlation functions by $ G^\prime({\bf r,r^\prime},T)=\theta(T) \, G_>^\prime({\bf r,r^\prime},T)+\theta(-T) \, G_<^\prime({\bf r,r^\prime},T) $. The prime denotes the particular choice of the gauge. Both $G_>^\prime$ and $G_<^\prime$ satisfy the homogeneous counterpart of Eq.(\ref{3114}). According to Horing \cite{Horing}, for arbitrarily chosen gauge they can be written as,    
\begin{equation}
\label{3134} G_{\{ {> \atop <} \}}({\bf r,r^\prime}, T )  =     \int \frac{d\omega}{2\pi} \, {\Bigg\{} {-i[1-f_0(\omega)]   \atop if_0(\omega)} {\Bigg\}}  \, \exp(-i\omega T)  \int_{-\infty}^\infty dT^\prime \, \exp(i\omega T^\prime) \, A({\bf r},{\bf r}^\prime,T^\prime)  
\end{equation}
with
\begin{eqnarray}
\label{3135}  A({\bf r},{\bf r}^\prime,T^\prime) & = & C({\bf r,r^\prime}) \int \frac{d{\bf p}}{(2\pi)^3} \, \exp(i{\bf p R}) \, \exp\left[ -i\left(\mu_B B\sigma_z+\frac{p_z^2}{2m}\right) T^\prime \right]  \nonumber\\
\label{3136} & & \times \frac{1}{\cos\left(\frac{\omega_c}{2} T^\prime\right)} \, \exp\left[ -i \frac{p_x^2+p_y^2}{m\omega_c} \tan\left(\frac{\omega_c}{2}T^\prime\right) \right] \, .
\end{eqnarray}
The gauge dependence of the Green's function is explicitly given in the factor $ C({\bf r,r^\prime}) $. Noting that $ C({\bf r,r^\prime}) $ is only a function of $ {\bf R}={\bf r} - {\bf r^\prime} $ and that it obeys the relation $ C({\bf r,r^\prime}) C({\bf r^\prime,r}) = 1$, this factor can be left aside in the following calculations. 
\subsection[Hartree-Fock term]{Hartree-Fock term}
First we calculate the Hartree-Fock term, which can be written in space time representation as  
\begin{equation}
\label{2.A.1}  \beta p_{HF} =  - \frac{1}{2\Omega} \, \sum_{ab} \int_0^1 \frac{d\lambda}{\lambda} \, {\bf Tr}_{(\sigma)} \int_0 \, d1 \, d2 \, V(12) \, G_a^\sigma(12) \, G_b^\sigma(21^+) \, \delta_{ab} \, .
\end{equation}
The free particle Green's function $ G_a^\sigma(12)$ must now be replaced by Eq.(\ref{3134}). In the resulting expression all integrals can be computed exactly. The detailed calculation is given in Appendix A. Defining $\xi_{ab}=e_a e_b/(4\pi \epsilon_0 kT \lambda_{ab})$ and $\lambda_{ab}=\hbar/\sqrt{2m_{ab} kT}$, $m_{ab}$ being the effectiv mass, we obtain the result
\begin{equation}
\label{2.A.2} \beta p_{HF} =  \sum_a \frac{\pi}{2} \, \tilde{z_a}^2 \, \lambda^3_{aa} \, \xi_{aa} \, f_1(x_a) \, , 
\end{equation}
where we have introduced
\begin{equation}
f_1(x_a) = \frac{\tanh(x_a)} {x_a}  \, \frac{ \cosh (2x_a)}{\cosh^2 (x_a) } \, \frac{{\rm arctanh}\sqrt{1-\frac{\tanh(x_a)}{x_a}} } {\sqrt{1-\frac{\tanh(x_a)}{x_a}}} \, .
\end{equation}

\subsection[Montroll-Ward term]{Montroll-Ward term}
Next we investigate the direct term of order $e^4$ given by the following expression 
\begin{equation}
\label{2.B.1}  \beta p_{MW} = \frac{1}{2 \Omega} \sum_{ab} \int_0^1 \frac{d\lambda}{\lambda} \, {\bf Tr_{(\sigma,\sigma^\prime)}} \int d1 \, d2 \, d3 \, d4 \, V_{ab}^s (12) \, V_{ab} (34) \,  G_a^\sigma(23) \, G_a^\sigma(32) \, G_b^{\sigma^\prime}(14) \, G_b^{\sigma^\prime}(41) \, .
\end{equation}
Again, a detailed calculation may be found in Appendix B. Retaining only contributions of order ${\tilde z}^2$ we obtain the result
\begin{equation}
\label{mwt88}  \beta p_{MW} =  \frac{\kappa^3}{12\pi}  -    \sum_{ab} \frac{\pi^\frac{3}{2}}{4} \, {\tilde z}_a{\tilde z}_b \, \lambda_{ab}^3 \, \xi_{ab}^2 \, f_2(x_a,x_b)  \, , 
\end{equation}
where $f_2(x_a,x_b)$ may be written as,
\begin{equation}
\label{mwt9}  f_2(x_a,x_b) = \left( \frac{1}{2} + \frac{4}{\pi} \int_0^1 dt \, \sqrt{t(1-t)} \, (y_a+y_b) \, \frac{{\rm arctanh}{\sqrt{1-(y_a+y_b)}}}{\sqrt{1-(y_a+y_b)}}\right) \, ,
\end{equation}
with $ y_{a,b} = \lambda_{aa,bb}^2 \, \sinh(x_{a,b}t) \, \sinh(x_{a,b}(1-t))/(\lambda_{ab}^2 \, t(1-t) \, 2 x_{a,b} \, \sinh(x_{a,b})) $. The first term in Eq.(\ref{mwt88}) is the Debye limiting law, while the second term gives a quantum correction. According to the Bohr-van-Leeuwen theorem the classical Debye law is not influenced by a magnetic field.

\subsection[second order exchange  term]{second order exchange term}
The exchange term of order $e^4$ is given by
\begin{equation}
\label{2.C.1.23} \beta p_{e^4} =   - \frac{1}{2\Omega} \sum_{ab} \int_0^1 \frac{d\lambda}{\lambda} \, {\bf Tr}_{{(\sigma)}} \int d1 \, d2 \, d3 \, d4 \, V_{ab} (13) \, V_{ab} (24) \,  G^\sigma(12) \, G^\sigma(23) \, G^{\sigma}(34) \, G^{\sigma}(41) \, \delta_{ab} \, .
\end{equation}
The result can be written in the form (Appendix C)
\begin{equation}
\label{2.C.1} \beta p_{e^4} =  - \sum_a \frac{\pi^{\frac{3}{2}} \ln{(2)}}{4} \, \, {\tilde z_a}^2 \, \lambda_{aa}^3 \, \xi_{aa}^2  \, f_3(x_a) \, ,
\end{equation}
where $f_3(x_a)$ is given by an integral representation (\ref{e4ap18}) and can only be evaluated numerically. Therefore we propose the following fit expression for $f_3(x_a)$
\begin{equation}
f_3(x_a) = \frac{\cosh(2x_a)}{\cosh^2(x_a)} \left( \frac{\tanh{(cx_a)}}{(cx_a)} \right)^{d} \, \frac{{\rm arctanh}{\sqrt{1-\frac{\tanh(cx_a)}{(cx_a)}}}}{\sqrt{1-\frac{\tanh(cx_a)}{(cx_a)}}} \, ,
\end{equation}
with the fitting parameters c= 0.8349 and d=0.9169. \par
\vspace{0.2cm}
Finally, we may sum up all contributions up to the order ${\tilde z}^2 e^4$. Collecting the obtained results (\ref{2.A.2},\ref{mwt88},\ref{2.C.1}), the scattering states contribution to the pressure in this approximation may be written as,    
\begin{equation}
\label{2.d.1} \beta p_{int}^{scatt}=\frac{\kappa^3}{12\pi} +\sum_{ab} {\tilde z}_a {\tilde z}_a B_{ab}^{scatt} \, ,
\end{equation}
where we have defined $B_{ab}^{scatt}$ by
\begin{equation}
\label{2.d.2} B_{ab}^{scatt}= \left( \delta_{ab} \, \frac{\pi}{2} \, \lambda_{ab}^3 \, \xi_{ab} \, f_1(x_a)   -  \frac{\pi^\frac{3}{2}}{4} \, \lambda_{ab}^3 \, \xi_{ab}^2 \, f_2(x_a,x_b) - \delta_{ab} \, \frac{\pi^{\frac{3}{2}}}{4} \, \ln{(2)} \,  \lambda_{ab}^3 \, \xi_{ab}^2 f_3(x_a)\right) \, .
\end{equation}

The influence of these states on the thermodynamics will be studied in section V and VI. Finally, we note that this equation gives in the limit $x_a \rightarrow 0$ the exact zero magnetic field results (see \cite{EKK}).

\section[Bound state contribution]{Bound state contribution}

According to Eq.(\ref{nd8}) we have for the bound state contribution 
\begin{equation}
\label{2.1.1} \beta p_{int}^{bound}=  z_e z_i P_3\sum_m e^{-\beta E_m} \, ,
\end{equation}
where $E_m$ are the eigenvalues of $\widehat H_{ab}^{\lambda=1}$. In Eq.(\ref{2.1.1}) all terms up to the order $e^4$ with respect to the interaction parameter must be omitted. In order to calculate  $p_{int}^{bound}$ the precise knowledge of the binding energies is essential. Therefore we briefly review the energy spectrum of the bound states and specify the approximations used in this paper. In contrast to the field-free hydrogen atom there is no exact solution for the nonrelativistic hydrogen atom at abritrary magnetic field strength. We focus  on the astrophysical interesting strong field regime $\gamma \gg 1$. Here we essentially follow the work of Lai\&Salpeter \cite{Lai&Salpeter}.\par
The two-body problem  has been investigated in the pseudomomentum approach \cite{Lai&Salpeter,Wunner1,Potekhin}. The pseudomomentum ${\bf K}=\sum_a ({\bf p_a}-e_a{\bf A_a}+e_a {\bf B}\times {\bf r_a})$ is a constant of motion. Therefore one can construct a wave function with a well-defined value of ${\bf K}$ by
\begin{equation}
\label{2.1.2} \psi ({\bf R},{\bf r})=\exp[i({\bf K}+(1/2){\bf B}\times {\bf r}) {\bf R}] \phi({\bf r}) \, , 
\end{equation}
with the centre of mass-coordinates ${\bf R}=(m_1{\bf r_1}+m_2{\bf r_2})/(m_1+m_2)$ and the relative coordinates ${\bf r}={\bf r_2}-{\bf r_1}$. Then the Hamiltonian of the Schr\"odinger equation $\widehat H \phi({\bf r})=(\widehat H_1+\widehat H_2) \phi({\bf r})=E_{nm\nu K_z K_\bot} \phi({\bf r})$ can be written in the form (setting ${\bf A}=1/2 \, ({\bf B} \times {\bf r})$ and $M=m_e+m_i$)
\begin{eqnarray}
\label{2.1.2.2} \widehat H_1 & = & \frac{{\bf p}^2}{2 m_{ei}}+\frac{e^2}{8m_{ei}} ({\bf B} \times{\bf r})^2+\left(\frac{1}{m_e}-\frac{1}{m_i} \right) \frac{e}{2} {\bf B} ({\bf r}\times {\bf p})-\frac{e^2}{4\pi \epsilon_0 r} \, , \\
\label{2.1.3} \widehat H_2 & = & \left(1+\frac{m_e}{m_i}\right) \frac{\hbar \omega_c^e}{2} + \frac{K_z^2}{2M}+\frac{{\bf K}_\bot^2}{2M}+\frac{e}{M} ({\bf K}\times {\bf B}){\bf r} \, .
\end{eqnarray}
In this approach the spectrum is characterized by the Landau quantum number n of the electron, the magnetic quantum number m, the number of nodes ${\nu}$ of the z wave function, and the pseudomomentum ${\bf K}$. In case $\gamma \gg 1$ we can restrict ourselves to n=0. The energy eigenvalues read as \cite{Lai&Salpeter}  
\begin{equation}
\label{3.1} E_{0m\nu K_z K_\bot}=E_{m\nu}+m \hbar \omega_c^e \frac{m_e}{m_i}+ \frac{K_z^2}{2M}+\frac{K_\bot^2}{2M_\bot}.    
\end{equation}
$E_{m\nu}$ is the energy of a bound electron moving in a fixed Coulomb potential. For $\nu=0$ the states are tightly bound with binding energies
\begin{equation}
\label{3.3} E_{m0}  =  -0.32 \, \frac{m_{ei}}{m_e} \, \ln^2\left(\frac{\gamma}{2m+1} \frac{m_e^2}{m_{ei}^2}\right) \, Ry \, ,
\end{equation}
while for $\nu\ge 1$ the states are hydrogen-like and the eigenvalues are well approximated by 
\begin{equation}
\label{3.4} E_{m\nu}=-\frac{1}{\nu_1^2} \, \frac{m_{ei}}{m_e} \, Ry \hspace{1cm} \nu_1=1,2,3,4...
\end{equation}
for the odd states (i.e. $\nu=2\nu_1-1$) and for the even states (i.e. $\nu=2\nu_1$). The second term in Eq.(\ref{3.1}) describes a Landau excitation of the proton which is coupled to the electron quantum number m due to the conservation of total pseudomomentum. The atom can freely move along the magnetic field direction contributing the term $K_z^2/2M$ to the energy. Contrary to that the transverse motion is coupled to the internal motion by the term $(e/M) ({\bf K} \times{\bf B} ) \, {\bf r}$. For magnetic field strengths considered here energy corrections due to this term can be computed by pertubation expansion with respect to the eigenstates of $\widehat H_1$. Lai\&Salpeter proposed an effective mass $M_\bot $ approximation of the transverse moving atom with 
\begin{equation} 
\label{3.5} M_{\bot} =M \left(1+ t \frac{\gamma}{0.32 \frac{M}{m_e} \ln\left(\gamma\right)}\right) \, , \hspace{1.5cm} t\approx 2.8 \, ,
\end{equation}
which we will use for simplification for all m-states. This energy correction is only valid for small pseudomomentum $K_\bot \ll K_{\bot c}$ where $K_{\bot c}$ is defined by $\hbar^2 K_{\bot c}^2/(2M) \approx (0.32 (M/m_e) \ln(\gamma)/(t\gamma)) \, Ry $  but serves as a fair approximation for magnetic fields strengths $B < 2.35\times 10^{9} T$.  
We note that due to the coupling of the intrinsic magnetic moment of the proton with the magnetic field an additional factor of $\left( 1+e^{-2x_i} \right)$ arises in the bound state partition function. On the other hand, at magnetic fields $\gamma \gg 1$ and temperatures $T \approx 10^{5-6}$ K spin excitations of the electrons can be neglected. \par
Given the energy eigenvalues we can define a convergent expression for the atomic partition function. The operator $P_3$ can be taken into account by subtracting the lowest order contributions with respect to the interaction parameter. As in the zero magnetic field case \cite{EKK} one can define a Planck-Larkin partition function    
\begin{equation} 
\label{3.2} \sigma_B(T) = \left( exp(-\beta E_{m0})-1\right) + \sum_{\nu=1} 2 \, \left( exp(-\beta E_{m\nu})-1+\beta E_{m\nu} \right) \, .
\end{equation}
Here, the factor 2 has its origin in the near-degeneracy of the hydrogen-like eigenstates. One can simplify the results by integrating over the pseudomomentum ${\bf K}$
\begin{equation}
\label{3.200} \int dK_z dK_\bot \exp{\left(-\frac{\beta K_z^2}{2M} - \frac{\beta K_\bot^2}{2M_\bot} \right)} = (2\pi M kT)^{\frac{3}{2}} \frac{M_\bot}{M} \, .  
\end{equation}

Now we can rewrite Eq.(\ref{2.1.1}). By using the eigenvalues $E_{0m\nu K_z K_\bot}$ (\ref{3.1}) and by introducing the modified fugacities ${\tilde z}_{e,i}$ according to Eq.(\ref{nd6}) we arrive at the following expression for the bound state contribution to the second virial coefficient
\begin{equation} 
\label{3.7} \beta p_{int}^{bound}= {\tilde z}_e \, {\tilde z}_i \, B_{ei}^{bound}={\tilde z}_e \, {\tilde z}_i \, 2 \, \pi^{\frac{3}{2}} \, \lambda_{ei}^3 \, \frac{\tanh{(x_e)}}{x_e} \frac{\tanh{(x_i)}}{x_i} \left( 1+e^{-2x_i} \right) {\sum_{m=0}} e^{-2mx_i} \, \frac{M_{\bot}}{M} \, \sigma_B(T) \, .
\end{equation}
$M_{\bot }$ and $\sigma_B(T)$ are given by Eq.(\ref{3.5}) and Eq.(\ref{3.2}), the energy eigenvalues $E_{m\nu}$ by Eq.(\ref{3.3}) and Eq.(\ref{3.4}), respectively.

\section[Equation of state]{Equation of state}

Now we can sum up all contributions we have considered. According to Eqs.(\ref{2.A.2},\ref{mwt88},\ref{2.C.1},\ref{3.7}) and expanding the ideal contribution in terms of the modified fugacities up to the order ${\tilde z}^2$ the pressure reads as follows
\begin{eqnarray}
\label{4.1} \beta p= \sum_a {\tilde z}_a + \frac{\kappa^3}{12\pi} + \sum_{ab} {\tilde z}_a {\tilde z}_b \, \bigg( -  \delta_{ab} \, \lambda_{ab}^3 \frac{\pi^\frac{3}{2}}{4} \frac{\tanh(x_a)}{x_a} \frac{\cosh(2 x_a)}{\cosh^2(x_a)} + \delta_{ab} \, \frac{\pi}{2} \, \lambda_{ab}^3 \, \xi_{ab} \, f_1(x_a)   -  \frac{\pi^\frac{3}{2}}{4} \, \lambda_{ab}^3 \, \xi_{ab}^2 \, f_2(x_a,x_b) \nonumber\\
\label{4.2} - \delta_{ab} \, \frac{\pi^{\frac{3}{2}}}{4} \, \ln{(2)} \,  \lambda_{ab}^3 \, \xi_{ab}^2 f_3(x_a)\bigg) +  {\tilde z}_e \, {\tilde z}_i \, B_{ei}^{bound} \, . 
\end{eqnarray}
The chemical potential in Eq.(\ref{4.2}) can be eliminated by using the relation
\begin{equation}
\label{4.21}  n_{e,i} = {\tilde z}_{e,i} \frac{\partial (\beta p)}{\partial {\tilde z}_{e,i}}
\end{equation}
to obtain the equation of state for a magnetized plasma. This procedure has been carried out numerically and the results are given in Fig. 1. 
\begin{figure}[h]
\centerline{\epsfig {figure=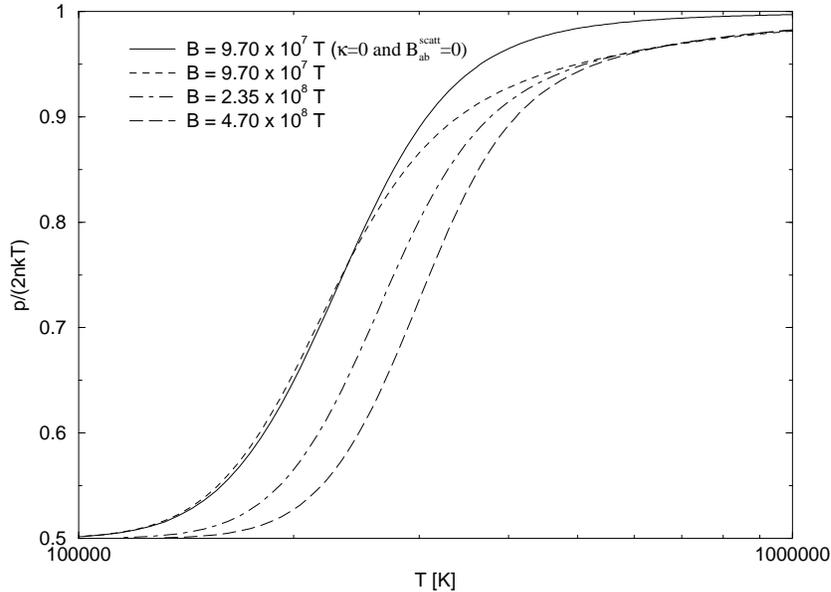,width=8.0cm,angle=-90}}
\caption{\sl The pressure for various magnetic field strengths at the density $n=10^{29} \, m^{-3} $ is plotted. For comparison the pressure without nonideality effects, i.e. $\kappa=0$ and $B_{ab}^{scatt}=0$, is shown. }
\end{figure}
Eqs.(\ref{4.2},\ref{4.21}) describe the ionization equilibrium of a weakly coupled hydrogen plasma in strong magnetic fields in an implicite form. The effect of the nonideality (i.e. of the scattering states contribution) of the plasma is to reduce the pressure. This contribution dominates the bound states contribution at high temperatures while at low temperatures the bound state term is dominant. 
Independent of the nonideality of the system we may also characterize the pressure by the magnetic field strength. For $T<6\times 10^5 K$ the pressure decreases with increasing magnetic field strength, while for $T>6\times 10^5 K$ the pressure increases as the magnetic field increases. This can be explained by the domination of the lowering of the ground state energy with increasing magnetic field strength at low temperatures, while at high temperatures the decrease of the phase space volume dominates. \par
In order to give a more explicit representation of the ionization equilibrium we will derive a generalized Saha equation in the next section. 

\section[Saha equation]{Saha equation}

In previous treatments of this problem \cite{Lai&Salpeter,Khersonskii,Miller} the interaction between the charged particles has been neglected. But at high densities considered here interactions between the particles play an important role. Our method is based on the chemical picture in which bound states are considered as composite particles, which must be treated on the same footing as elementary particles. By inspection of the fugacity expansion (\ref{4.1}) we reinterpret the term containing the partition function $\sigma_B(T)$ as the fugacity $z_0^\star$ of the neutral atoms

\begin{equation} 
\label{5.1} {\tilde z}_0^\star={\tilde z}_i {\tilde z}_e B_{ei}^{bound} \, .
\end{equation}
Defining the fugacities of the free composite particles in the chemical picture by ${\tilde z}_e^\star={\tilde z}_e \, , \, {\tilde z}_i^\star={\tilde z}_i$ the pressure reads as follows
\begin{equation} 
\label{5.2} \beta p={\tilde z}_e^\star+{\tilde z}_i^\star+\frac{\kappa^{\star 3}}{12\pi}+\sum_{ab} {\tilde z}_a^\star {\tilde z}_b^\star B_{ab}^{free}+{\tilde z}_0^\star \, ,
\end{equation}
with $B_{ab}^{free}=B_{ab}^{scatt}+B_{ab}^{ideal}$ and $ B_{ab}^{ideal} $ is given by
\begin{equation} 
\label{5.2.1} B_{ab}^{ideal} = -  \delta_{ab} \, \lambda_{ab}^3 \frac{\pi^\frac{3}{2}}{4} \frac{\tanh(x_a)}{x_a} \frac{\cosh(2 x_a)}{\cosh^2(x_a)} \, .
\end{equation}
The particle densities of the new species are given by
\begin{equation} 
\label{5.3} n_e^\star={\tilde z}_e^\star \frac{\partial \left( \beta p\right)}{\partial {\tilde z}_e^\star} \, , \hspace{1cm}  n_i^\star={\tilde z}_i^\star \frac{\partial \left( \beta p\right)}{\partial {\tilde z}_i^\star} \, , \hspace{1cm} n_0^\star={\tilde z}_0^\star \frac{\partial \left( \beta p\right)}{\partial {\tilde z}_0^\star} \, .
\end{equation}
Solving this equation by iteration we find 
\begin{eqnarray} 
\label{5.4} \ln {\tilde z}_e^\star & = & \ln n_e^\star -\frac{1}{2} \frac{\beta e^2 \kappa^\star}{4\pi \epsilon_0}-2n_e^\star (B_{ee}^{free}+B_{ei}^{free}) \, , \nonumber\\
\label{5.5} \ln {\tilde z}_i^\star & = & \ln n_i^\star -\frac{1}{2} \frac{\beta e^2 \kappa^\star}{4\pi \epsilon_0}-2n_i^\star (B_{ii}^{free}+B_{ei}^{free}) \, , \nonumber\\
\label{5.6}  \ln {\tilde z}_0^\star & = & \ln n_0^\star \, ,
\end{eqnarray}
where now $\kappa^{\star 2}=(n_e^\star+n_i^\star) \beta e^2/\epsilon_0=2 n_e^\star \beta e^2/\epsilon_0$. By inserting the fugacities according to Eq.(\ref{5.6}) into Eq.(\ref{5.1}) the following Saha equation is obtained
\begin{equation} 
\label{5.7} \frac{n_0^\star}{n_e^\star n_i^\star}= B_{ei}^{bound} \, \exp\left(-\frac{\beta e^2 \kappa^\star}{4 \pi \epsilon_0} - 2 n_e^\star \sum_{ab} B_{ab}^{free}\right) \, ,
\end{equation}
where $ B_{ab}^{free} $ is to be taken from Eq.(\ref{2.d.2}) and (\ref{5.2.1}). It is useful to extend the range of validity of Eq.(\ref{5.7}) for large $\xi_{ab}$ by a kind of Pad\'{e} approximation. Noting that 
\begin{equation} 
-\frac{\beta e^2 \kappa^\star}{4 \pi \epsilon_0} - 2 n_e^\star \sum_{ab} B_{ab}^{free}=-\frac{\beta e^2 \kappa^\star}{4 \pi \epsilon_0} \left(1- \kappa^\star a\right) \approx - \frac{\beta e^2 \kappa^\star}{4 \pi \epsilon_0} \frac{1}{\left(1+\kappa^\star a\right)}  \, ,
\end{equation}
where $a$ may interpreted as an effective radius of the charged particles and is defined by
\begin{eqnarray} 
\label{5.9.1.2} a=  \frac{4\pi \epsilon_0^2}{\beta^2 e^4} \sum_{ab} & {\Bigg(}   &   \frac{\pi^\frac{3}{2}}{4} \, \lambda_{ab}^3 \, \xi_{ab}^2 \, f_2(x_a,x_b) + \delta_{ab} \, \frac{\pi^{\frac{3}{2}}}{4} \, \ln{(2)} \,  \lambda_{ab}^3 \, \xi_{ab}^2 f_3(x_a) - \delta_{ab} \, \frac{\pi}{2} \, \lambda_{ab}^3 \, \xi_{ab} \, f_1(x_a)    \nonumber\\
\label{5.9.1} & + & \delta_{ab} \, \lambda_{ab}^3 \frac{\pi^\frac{3}{2}}{4} \frac{\tanh(x_a)}{x_a} \frac{\cosh(2 x_a)}{\cosh^2(x_a)}  {\Bigg)} \approx \frac{\sqrt{\pi}}{16} \sum_{ab} \lambda_{ab} \left(f_2(x_a,x_b)+\ln(2) f_3(x_a)\right)\, ,
\end{eqnarray}
we find the modified Saha equation
\begin{equation} 
\label{5.8} \frac{n_0^\star}{n_e^\star n_i^\star}= B_{ei}^{bound} \, \exp\left(-\frac{\beta e^2 \kappa^\star}{4 \pi \epsilon_0 (1+\kappa^\star a)}\right) \, .
\end{equation}
The Eq.(\ref{5.8}) differs from the Saha equation given in \cite{Lai&Salpeter} by an additional exponential factor which may be interpreted as the lowering of the ionization energy. In Fig. 2 the degree of ionization $\alpha=n_e^\star/n$ for a dense hydrogen plasma at various magnetic field strengths is plotted and compared with the results of the ideal Saha equation \cite{Lai&Salpeter}. We find an increase of the ionization degree in comparison with the ideal Saha equation \cite{Lai&Salpeter} due to the nonideality effects. For densities of about $10^{29}-10^{30} m^3$ the deviation from the ideal Saha equation may be as large as 10-15\% (see Fig. 2). At even higher densities, i.e., $n\gg 10^{30} m^{-3}$, this result may only be used as a rough approximation. The plasma can no longer be regarded as a weakly coupled system rather it must be treated as a strongly coupled system. \par
Additionally, we may characterize the dependence of the ionization degree on the magnetic field strength. With increasing magnetic field strength the ionization degree decreases at temperatures $T<6\times 10^5 K$, while for temperatures $T>6\times 10^5 K$ the ionization degree increases. The explanation of this effect was given in section V.

\begin{figure}[h]
\centerline{\epsfig {figure=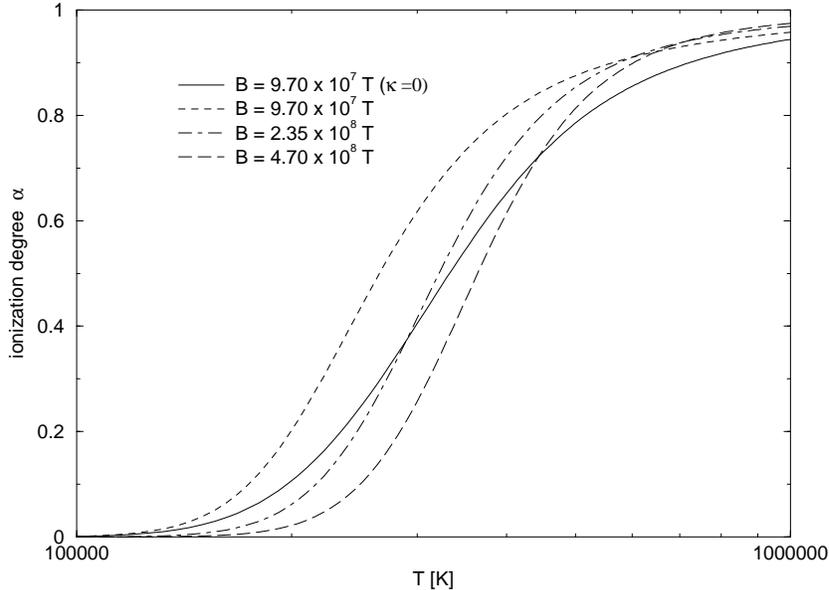,width=8.0cm,angle=-90}}
\caption{\sl Degree of ionization at a density of $\rho \approx 2 \, g \, cm^{-3} (n=1\times 10^{30} m^{-3})$ for various magnetic field strength. The ionization fraction for $\kappa=0$ is included (solid line).}
\end{figure}

\section[Conclusion]{Conclusion}
In this paper we constructed a theory describing a hydrogen plasma in a constant uniform magnetic field. Starting from a fugacity expansion we derived a general expression for the second virial coefficient as a perturbation expansion with respect to the interaction parameter $e^2$ and we explicitly calculated the lowest order contributions for the scattering part and considered bound state contributions at arbitrary order by using the approximate results for the binding energy of Lai\&Salpeter \cite{Lai&Salpeter}. The results were used to establish the equation of state. Finally, we have derived a generalized Saha equation and we have shown that at high densities and at temperatures typical for the surface of neutron stars nonideality effects can significantly increase the degree of ionization. \par
The accuracy of the absolute values of the considered physical quantities can be improved by using more accurate energy eigenvalues, i.e. better fitting formulas, and by calculating even higher order contributions to the scattering part of the second virial coefficient. Nevertheless, the influence of the nonideality effects on the ionization equilibrium as shown in this paper remains approximately the same.

\section[Acknowledgments]{Acknowledgments}
This work was supported by the Deutsche Forschungsgemeinschaft under grant\#Eb 126/5-1.

\begin{appendix}
\section[Hartree-Fock term]{Hartree-Fock term}
By using the representation of the Green's function in terms of the spectral function (Eq.(\ref{3136})) we obtain for the Hartree-Fock term 
\begin{eqnarray}
\label{eq229}  \langle V \rangle_{HF} & = &  \frac{1}{2} \sum_a \, {\bf Tr}_{(\sigma)} \int \frac{d{\bf p}_1}{(2\pi)^3} \int \frac{d{\bf p}_2}{(2\pi)^3} \, \frac{e_a^2}{\epsilon_0 \mid {\bf p}_1-{\bf p}_2 \mid ^2} \int d\omega_1 \, f_0(\omega_1) \int d\omega_2 \, f_0(\omega_2) \,  \int dT_1 \int dT_1 \nonumber\\
\label{eq230} && \times \, e^{i\omega_1 T_1} \, e^{i\omega_2 T_2}  A_a^\sigma({\bf p}_1,T_1) \, A_a^\sigma({\bf p}_2,T_2) \, .
\end{eqnarray}
In order to fulfill the periodicity condition of the Green's function every time variable must be extended in the complex time region. Therefore we associate with each time variable a small negative imaginary part $t \rightarrow t(1-i\delta)$ and the corresponding integration may be taken in the sense of an inverse Laplace transform. Inserting Eq.(\ref{3136}) , the Hartree-Fock pressure should be written as,  
\begin{eqnarray}
\label{HFND1}  \langle V \rangle_{HF} &  = &   \frac{1}{2}  \sum_a \int {\frac{d\bf p_1}{(2\pi)^3}} \int \frac{d{\bf p}_2}{(2\pi)^3} \, \frac{ e_a^2}{\epsilon_0 \mid {\bf p}_1-{\bf p}_2 \mid ^2} \int d\omega_1 \, f_0(\omega_1) \int d\omega_2 \, f_0(\omega_2) \int_{\delta - i\infty}^{\delta + i\infty} \frac{ds_1}{2\pi i}  \, e^{\omega_1 s_1}  \nonumber\\
\label{HFND2} & & \times \int_{\delta-i\infty}^{\delta + i\infty}   \frac{ds_2}{2\pi i} \hspace{0.1cm}\, e^{\omega_2 s_2} \, \frac{2 \cosh \left(\frac{\omega_c^a}{2}(s_1+s_2)\right)}{\cosh \left(\frac{\omega_c^a}{2}s_1\right) \cosh \left(\frac{\omega_c^a}{2}s_2\right)} \, \exp\left(-\frac{p_{1z}^2}{2m_a} s_1\right) \, \exp\left(-\frac{p_{2z}^2}{2m_a} s_2\right)\nonumber\\
\label{HFND3} & & \times \exp\left(-\frac{p_{1\rho}^2}{m_a\omega_c^a} \tanh{\left(\frac{\omega_c^a}{2} s_1\right)}\right) \, \exp\left(-\frac{p_{2\rho}^2}{m_a\omega_c^a} \tanh{\left(\frac{\omega_c^a}{2} s_2\right)}\right) \hspace{0.12cm}.
\end{eqnarray}
This integral may be simplified in the nondegenerate case, $f_0(\omega) \rightarrow e^{\beta \mu} e^{-\beta \omega}$, where the $\omega$ and s integration are Laplace transform and inverse, so that    
\begin{eqnarray}
\label{HFND4}  \langle V \rangle_{HF} & = &  \sum_a \, z_a^2 \int {\frac{d{\bf p}_1}{(2\pi)^3}} \int \frac{d{\bf p}_2}{(2\pi)^3} \, \frac{ e_a^2}{\epsilon_0 \mid {\bf p}_1 - {\bf p}_2 \mid ^2}  \,  \frac{\cosh \left(\omega_c^a \beta\right)}{\cosh^2 \left(\frac{\omega_c^a}{2} \beta\right) } \, \exp\left( -\frac{p_{1z}^2+p_{2z}^2}{2m_a}\beta \right) \nonumber\\
\label{HFND5} & &  \times \exp\left(-\frac{p_{1x}^2+p_{1y}^2+p_{2x}^2+p_{2y}^2}{m_a\omega_c^a} \tanh \left(\frac{\omega_c^a}{2}\beta \right)\right) \, .
\end{eqnarray} 
The Gaussian momentum integrations are readily carried out, with the result
\begin{eqnarray}
\label{HFND12} \langle V \rangle_{HF} & = &   \sum_a \, \tilde{z}_a^2 \, \lambda_{aa}^2 \, \frac{\tanh(x_a)} {x_a} \, \frac{\pi^{\frac{3}{2}}}{2^{\frac{3}{2}}} \int {\frac{d{\bf p}_1}{(2\pi)^3}} \, \frac{e_a^2}{\epsilon_0 \mid {\bf p}_1\mid ^2}   \, \frac{\cosh (2x_a)}{\cosh^2 (x_a) } \, \exp\left( -\frac{p_{1z}^2}{2}  \right) \nonumber\\
\label{HFND13} & &  \times \exp\left(-\frac{p_{1x}^2+p_{1y}^2}{2x_a} \tanh(x_a) \right)  \, .
\end{eqnarray}
The remaining integrals with respect to ${\bf p}_1$ may be evaluated exactly and the result can be expressed in terms of elementary functions \cite{Gradstein&Ryshik}
\begin{equation}
\label{HFND17} \langle V \rangle_{HF}  =   \sum_a \,  \frac{\pi}{2} \, \tilde{z}_a^2 \, \lambda_{aa}^2 \, \frac{e_a^2}{4 \pi \epsilon_0} \, \frac{\tanh(x_a)} {x_a}  \, \frac{ \cosh (2x_a)}{\cosh^2 (x_a) }  \, \frac{arctanh\sqrt{1-\frac{\tanh(x_a)}{x_a}} } {\sqrt{1-\frac{\tanh(x_a)}{x_a}}}   \, .
\end{equation}
Finally, the charging integral may be carried out to obtain the Hartree-Fock contribution given in Eq.(\ref{2.A.2}).
\section[Montroll-Ward term]{Montroll-Ward term}
According to Eq.(\ref{2.B.1}), the Montroll-Ward term may be written as, 
\begin{eqnarray}
\label{ap2.3}  \langle V \rangle_{MW} & = & \frac{i}{2} \sum_{ab} {\bf Tr}_{(\sigma,\sigma^\prime)} \int \frac{d{\bf q}}{(2\pi)^3} \int \frac{d{\bf p}}{(2\pi)^3} \int \frac{d{\bf k}}{(2\pi)^3} \beta \, \int_0^{-i\beta} dt \, V_{ab}^s({\bf q}) \,  V_{ab}({\bf q}) \nonumber\\
\label{ap2.4}  & &    \times G_a^{\sigma >}({\bf p}-\frac{{\bf q}}{2};t) \, G_a^{\sigma <} ({\bf p}+\frac{{\bf q}}{2};-t) \, G_b^{\sigma^\prime >}({\bf k}-\frac{{\bf q}}{2};t) \, G_b^{\sigma^\prime <}({\bf k}+\frac{{\bf q}}{2};-t) \, . 
\end{eqnarray}
We are interested in the low density region, i.e. $f_0(\omega)<1$. Thus we consider only contributions up to the order $z^2$. Applying the same arguments as discussed in the previous section leads to the equation
\begin{eqnarray}
\label{ap2.12} \langle V \rangle_{MW}  & = &  \frac{i}{2} \sum_{ab} z_a z_b \, {\bf Tr}_{(\sigma,\sigma^\prime)} \int \frac{d{\bf q}}{(2\pi)^3} \int \frac{d{\bf p}}{(2\pi)^3} \int \frac{d{\bf k}}{(2\pi)^3} \hspace{0.1cm} \int_0^{-i\beta} dt \, V_{ab}^s({\bf q}) \, V_{ab}({\bf q}) \\
\label{ap2.13} && {\rm  \times   A_a^\sigma({\bf p-\frac{q}{2}} , t) \, A_a^\sigma({\bf p+\frac{q}{2}} , -i\beta - t) \, A_b^{\sigma^\prime}({\bf k-\frac{q}{2}} , t) \, A_b^{\sigma^\prime}({\bf k+\frac{q}{2}} , -i\beta - t) }      \, . \nonumber
\end{eqnarray}
Again, $A({\bf k})$ may be replaced according to Eq.(\ref{3136}) and all Gaussian integrals may be evaluated with the result   
\begin{equation}
\label{ap2.21}  \langle V \rangle_{MW}  =  \frac{\beta}{2} \sum_{ab} \frac{{\tilde z}_a {\tilde z}_b}{(2\pi)^3} \, \left(\frac{e_a e_b}{\epsilon_0}\right)^2 \, \lambda_{ab} \,  \int_0^1 dt \int d{\bf q} \, \frac{1}{{\bf q}^2+\kappa^2 \lambda_{ab}^2} \, \frac{1}{ {\bf q}^2} \exp\left(-q_z^2  \, t (1-t)\right) \, \exp\left(-q_\rho^2 \, t (1-t) \, (y_a+y_b)\right) \, , 
\end{equation}
where we have defined $ y_{a,b} = \lambda_{aa,bb}^2 \, \sinh(x_{a,b}t) \, \sinh(x_{a,b}(1-t))/(\lambda_{ab}^2 \, t(1-t) \, 2 x_{a,b} \, \sinh(x_{a,b})) $. Introducing spherical coordinates one can integrate with respect to $q$. The result is readily seen to be
\begin{eqnarray}
\label{ap2.22}  \langle V \rangle_{MW} & = & \frac{\beta}{2} \sum_{ab} \frac{ {\tilde z}_a {\tilde z}_b}{(2\pi)^3} \, \left(\frac{e_a e_b}{\epsilon_0}\right)^2  \,  \int_0^1 dt \, \frac{\pi^2}{\kappa} \int_{-1}^1 dz  \, \exp\left( \kappa^2 \lambda_{ab}^2 t(1-t) \left(y_a+y_b - z^2(y_a+y_b-1)\right) \right)  \nonumber\\
\label{ap2.22extra} && \times \left(1-erf\left(\kappa \lambda_{ab} \sqrt{t(1-t) (y_a+y_b - z^2 (y_a+y_b-1))} \right) \right) \, .
\end{eqnarray}
Finally, the z-integration yields, 
\begin{eqnarray}
\label{ap2.24}   \langle V \rangle_{MW} & = & \frac{\beta}{2} \sum_{ab} \frac{{\tilde z}_a {\tilde z}_b}{(2\pi)^3} \, \left(\frac{e_a e_b}{\epsilon_0}\right)^2  \, \int_0^1 dt \, \frac{\pi^2}{\kappa  }  \bigg( \frac{2\exp\left( \kappa^2 \lambda_{ab}^2 t(1-t) \left(y_a+y_b\right)\right)}{\kappa \lambda_{ab} \sqrt{t(1-t) ((y_a+y_b-1))}}   erf\left(\kappa \lambda_{ab} \sqrt{t(1-t) ((y_a+y_b-1))}\right)   \nonumber\\
\label{ap2.24extra} & &  - \frac{4}{\sqrt \pi}  \sum_{k=0}^\infty \, \frac{2^k \, \left(\kappa \lambda_{ab} \sqrt{t(1-t) ((y_a+y_b-1))} \right)^{2k+1}}{(2k+1)!!}  \, _2F_1\left( \frac{1}{2},-k-\frac{1}{2};\frac{3}{2}, 1-\frac{1}{y_a+y_b} \right)  \bigg) \, .
\end{eqnarray}
For a low density plasma we may expand this expression in powers of $\kappa \lambda$ and retain only contributions to first order. Using the representation of the hypergeometric function 
\begin{equation}
\label{mwt1} _2F_1\left( \frac{1}{2},-\frac{1}{2};\frac{3}{2}, x^2\right) = \frac{1}{2} \left(\sqrt{1-x^2}+\frac{\arcsin(x)}{x}\right)
\end{equation}
the Montroll-Ward contribution to the second virial coefficient becomes
\begin{equation}
\label{mwt8} \langle V \rangle_{MW}  =  kT \, \frac{\kappa^3}{8\pi}  -    \sum_{ab} \frac{\pi^\frac{3}{2}}{2} \, kT \, {\tilde z}_a{\tilde z}_b \, \lambda_{ab} \, \left(\frac{e_a e_b \beta}{4\pi \epsilon_0}\right)^2 \, \left( \frac{1}{2} + \frac{4}{\pi} \int_0^1 dt \, \sqrt{t(1-t)} \, (y_a+y_b) \, \frac{arctanh{\sqrt{1-(y_a+y_b)}}}{\sqrt{1-(y_a+y_b)}}\right)  \, .
\end{equation}
After performing the charging procedure one may obtain the Montroll-Ward contribution to the pressure (Eq.(\ref{mwt88})).
\section[second order Exchange term]{second order Exchange term}
This contribution is found to be
\begin{eqnarray}
\label{e4ap2} \langle V \rangle_{e^4} &  = & \frac{i}{2}  \sum_a {\bf Tr}_{(\sigma)} \int_0^{-i\beta} dt \int \frac{d{\bf p}}{(2\pi)^3} \int \frac{d{\bf q}}{(2\pi)^3} \int \frac{d{\bf k}}{(2\pi)^3} \, V({\bf q}) \, V({\bf k}) \, G_a^{\sigma >}({\bf p}+\frac{{\bf q}}{2} +\frac{{\bf k}}{2};t)  \nonumber\\
\label{e4ap3} & &  \times G_a^{\sigma <}({\bf p}-\frac{{\bf q}}{2} +\frac{{\bf k}}{2};-t) \, G_a^{\sigma >}({\bf p}-\frac{{\bf q}}{2} -\frac{{\bf k}}{2};t) \, G_a^{\sigma <}({\bf p}+\frac{{\bf q}}{2} -\frac{{\bf k}}{2};-t) \, ,
\end{eqnarray}
where the screened potential $V^s$ was replaced by the bare Coulomb potential $V$. Performing the Laplace transform and inverse, this equation may be rewritten as,
\begin{eqnarray}
\label{e4ap4} \langle V \rangle_{e^4} &  = &  \frac{i}{2}  \sum_a \, z_a^2 \, {\bf Tr}_{(\sigma)} \int_0^{-i\beta} dt \int \frac{d{\bf p}}{(2\pi)^3} \int \frac{d{\bf q}}{(2\pi)^3} \int \frac{d{\bf k}}{(2\pi)^3} \, V({\bf q}) \, V({\bf k}) \, A_a^{\sigma}({\bf p}+\frac{{\bf q}}{2} +\frac{{\bf k}}{2};t)  \nonumber\\
\label{e4ap5} & & \hspace{-0.26cm} \times A_a^{\sigma}({\bf p}-\frac{{\bf q}}{2} +\frac{{\bf k}}{2};-i\beta-t) \, A_a^{\sigma }({\bf p}-\frac{{\bf q}}{2} -\frac{{\bf k}}{2};t) \, A_a^{\sigma}({\bf p}+\frac{{\bf q}}{2} -\frac{{\bf k}}{2};-i\beta-t) \, . 
\end{eqnarray}
Carrying out all elementary integrals we obtain the result 
\begin{equation}
\label{e4ap17} \langle V \rangle_{e^4} = kT \sum_a \, \frac{\pi^{\frac{3}{2}} \ln{(2)}}{2} \, \lambda_{aa} \, \left(\frac{e_a^2 \beta}{4\pi \epsilon_0}\right)^2  \, {\tilde z}_a^2 \,  f_3(x_a) \, ,
\end{equation}  
where $f_3(x_a)$ is given by the integral representation 
\begin{equation}
\label{e4ap18}  f_3(x_a)   =   \frac{1}{\pi \ln{(2)}}  \, \frac{\cosh{(2x_a)}}{\cosh^2{(x_a)}} \int_0^1 dt \int_0^\infty dt_1  \, \frac{1}{\sqrt{t_1+4t(1-t)}} \frac{arctanh{\sqrt v_a}}{\sqrt v_a} \frac{1}{t_1 x_a/ [\tanh(x_a t)+\tanh(x_a (1-t))] +1}\, , 
\end{equation}
with
\begin{eqnarray}
\label{e4ap19}  v_a  =   1 & - & \frac{t_1 [\tanh(x_a t)+\tanh(x_a (1-t))]/x_a+4 [\tanh(x_a t) \tanh(x_a (1-t))]/x_a^2}{t_1+4 t (1-t)} \nonumber\\
& \times &  \frac{t_1+1}{t_1 +[\tanh{(x_at)}+\tanh{(x_a(1-t))}]/ x_a} \, .     
\end{eqnarray}
The charging procedure yields an additional factor $1/2$ and, finally, one obtains for $e^4$-exchange term the result given Eq.(\ref{2.C.1}).

\end{appendix}

\end{document}